\documentclass{article}

\usepackage[preprint]{ms}

\usepackage{natbib} 
\usepackage[utf8]{inputenc} 
\usepackage[T1]{fontenc}    
\usepackage{hyperref}       
\usepackage{url}            
\usepackage{booktabs}       
\usepackage{doi}            

\usepackage{amsfonts}       
\usepackage{nicefrac}       
\usepackage{microtype}      
\usepackage{xcolor}         

\usepackage{amsmath}
\usepackage{amssymb}
\usepackage{bbm}
\usepackage{mathtools}
\usepackage{amsthm}
\usepackage{amsfonts}
\usepackage{centernot}

\usepackage{caption}
\usepackage{subcaption}
\usepackage{bbm}
\usepackage{enumitem}

\title{Integrated nowcasting of convective precipitation with Transformer-based models using multi-source data}

\author{%
	\c{C}a\u{g}lar K\"u\c{c}\"uk\thanks{Corresponding author: \texttt{caglar.kucuk@geosphere.at}} 
	\hspace*{1cm}Aitor Atencia
	\hspace*{1cm}Markus Dabernig\\
	GeoSphere Austria -- Federal Institute for Geology, Geophysics, Climatology and Meteorology
}

\begin{document}
	
\maketitle
	
\begin{abstract}
Precipitation nowcasting is crucial for mitigating the impacts of severe weather events and supporting daily activities. Conventional models predominantly relying on radar data have limited performance in predicting cases with complex temporal features such as convection initiation, highlighting the need to integrate data from other sources for more comprehensive nowcasting. Unlike physics-based models, machine learning (ML)-based models offer promising solutions for efficiently integrating large volumes of diverse data. We present EF4INCA, a spatiotemporal Transformer model for precipitation nowcasting that integrates satellite- and ground-based observations with numerical weather prediction outputs. EF4INCA provides high-resolution forecasts over Austria, accurately predicting the location and shape of precipitation fields with a spatial resolution of 1 kilometre and a temporal resolution of 5 minutes, up to 90 minutes ahead. Our evaluation shows that EF4INCA outperforms conventional nowcasting models, including the operational model of Austria, particularly in scenarios with complex temporal features such as convective initiation and rapid weather changes. EF4INCA maintains higher accuracy in location forecasting but generates smoother fields at later prediction times compared to traditional models. Interpretation of our model showed that precipitation products and SEVIRI infrared channels CH7 and CH9 are the most important data streams. These results underscore the importance of combining data from different domains, including physics-based model products, with ML approaches. Our study highlights the robustness of EF4INCA and its potential for improved precipitation nowcasting. We provide access to our code repository, model weights, and the dataset curated for benchmarking, facilitating further development and application.
\end{abstract}
	
\section{Introduction}
Nowcasting, the short-term prediction of weather conditions within six hours, is crucial for mitigating the impacts of severe weather events. It plays a vital role in various domains, including daily activities such as aviation and renewable energy management, as well as extreme events that trigger civil protection operations \citep{WMO2021}. Traditional models for precipitation nowcasting primarily use radar observations, which are extrapolated to predict future time steps \citep{Wapler2019}. However, this approach has limited capacity in predicting convective development because it relies on Lagrangian extrapolation of the latest observations, which does not account for the initiation, growth, or decay of convection and precipitation. Overall, the performance of precipitation nowcasting models is hampered in cases with complex temporal features such as the onset of convection, which remains the largest source of error in nowcasting \citep{Prudden2020}, and the presence of fast moving or decaying fields. Despite numerous approaches to combine radar data with additional sources \citep[e.g., ][]{Nisi2014,Mecikalski2015,James2018,Cintineo2018,Cintineo2020}, success has been limited due to the underlying models not being explicitly designed to extract information from large volumes of data. 

As an alternative to conventional models, various machine learning (ML)-based approaches have been suggested for nowcasting. Similar to the initial developments in meteorology, early ML models were not designed to incorporate diverse data streams, despite providing impactful methods for the ML community. The problem of learning spatial and temporal connections simultaneously was initially addressed by combining convolutional and recurrent neural networks, facilitating the development of powerful methods such as Convolutional LSTMs \citep{Shi2015} and Trajectory GRUs \citep{Shi2017}. These methods were subsequently incorporated into generative models \cite[e.g.,][]{Ravuri2021} to prevent the loss of the sharp spatial structure of precipitation fields during predictions. Furthermore, Transformer-based models were proposed for Earth system science problems as an alternative to the conventional ML models built upon convolutional and recurrent neural networks \citep{Earthformer}.

Nevertheless, there have been successful efforts towards ML-based integrated nowcasting. \citet{Leinonen2023a} leveraged input data from various sources, such as satellite- and ground-based observations, alongside model data, through a combination of recurrent and convolutional neural networks to predict various hazardous weather phenomena such as lightning, hail, and heavy precipitation. Interpretation of their model demonstrated the importance of supporting radar data with other data streams, such as satellite observations and numerical weather prediction (NWP) products, particularly for longer lead times.

Another notable example of ML-based integrated nowcasting is the MetNet model family. The latest model in this family, MetNet-3 \citep{Andrychowicz2023}, utilises not only spatiotemporal data from diverse sources, but also point-based data through Vision Transformers in a U-Net architecture. Owing to the high performance of this model across the United States (US), it is currently used by Google as the default weather prediction app in that spatial domain. Despite its ability to predict precipitation, MetNet-3 is highly customised for input data in the US, with code and model weights unavailable, and very expensive to train.

In this paper, we present EF4INCA, a spatiotemporal Transformer model based on Earthformer \citep{Earthformer}, which is tailored for convective precipitation nowcasting that integrates data from various streams. EF4INCA predicts precipitation over Austria with a spatial resolution of 1 kilometre and a temporal resolution of 5 minutes for a lead time of up to 90 minutes. We describe the dataset created to train the model, the methodological details, and present model outputs through case studies. We compare predictions from EF4INCA against the operational precipitation nowcasting model over Austria  \cite[INCA;][]{haiden2011integrated} and Pysteps \citep{Pulkkinen2019} as a benchmark nowcasting framework, both in case studies and through prediction scores over the test dataset. We perform permutation tests to interpret EF4INCA, discuss the findings, and conclude with a summary and potential outlook for ML-based integrated nowcasting.
	
\section{Data}\label{sec:Data}
\subsection{Observation data}
\subsubsection{Satellite data}

Satellite-based observations are acquired from the Spinning Enhanced Visible and InfraRed Imager \cite[SEVIRI;][]{SEVIRI} sensor aboard the Meteosat Second Generation (MSG) satellite. Specifically, we used the Rapid Scan Service, which is available over Europe with 5-minute temporal resolution. We used four infrared channels with central wavelengths of $6.2$, $7.3$, $8.7$, and $10.8 \mu m$ (hereafter CH5, CH6, CH7, and CH9, respectively). 

CH5 and CH6 are commonly used in Numerical Weather Prediction (NWP) systems, as they provide information about the water vapour in the upper and lower troposphere, respectively. CH7 and CH9 are less sensitive to water vapour content but provide diagnostics about atmospheric features and cloud properties through brightness temperatures. The usefulness of CH9, in particular, has been demonstrated in various studies, ranging from purely meteorological \cite[e.g.,][]{Schmit2018} to data-driven approaches \cite[e.g.,][]{Kucuk2024}.

After downloading the level 1.5 product using the EUMETSAT Data Access Client\footnote{https://user.eumetsat.int/resources/user-guides/eumetsat-data-access-client-eumdac-guide}, we converted reflectance values into brightness temperatures, reprojected, and cropped the data to match the operational INCA domain. 

\subsubsection{Ground-based observations}
	
\paragraph{Precipitation radar}
We used precipitation observations from ground-based radar towers that are used in INCA predictions. The radars operate in C-band, providing a resolution of 5 minutes in time and 1 km in space \cite[see][for details]{Kaltenboeck2015}. 2D precipitation estimates were mosaicked from these towers over the study domain and used them as the data stream for ground-based precipitation observations.

\paragraph{Lightning} 
We used ground-based lightning observations collected within the European Cooperation for Lightning Detection network \cite[EUCLID;][]{EUCLID}. To ensure compatibility with satellite-based lightning data, 30 \% of the detected events were discarded based on their measured discharge amplitude, given the potential for a 30 \% miss rate of satellite-based lightning detection systems \citep{Goodman2013}. After filtering out 30 \% of the events with the lowest discharge amplitude, lightning occurrences per grid cell and time step were counted to generate gridded data. We stored these counts in tabular format in each sample to conserve storage, but converted them into raster on-the-fly while using the data.

\subsection{Modelled data}{\label{ssec:ModelData}}

We used Integrated Nowcasting through Comprehensive Analysis (INCA) model outputs for precipitation analysis, hereafter INCA\_a, and convective available potential energy (CAPE) estimates in our model. The rapid (5-minute) version of the INCA system, as described by \citet{haiden2011integrated}, provides multivariable analysis and nowcasting over an area of 400x700 km centred on Austria. It generates near-real-time analyses and forecasts for various meteorological fields, including precipitation, temperature, moisture, global radiation, and CAPE. The system is designed to enhance numerical forecast products within the nowcasting (0-4 hours) and very short (up to 12 hours) time frames. 

For estimation of INCA\_a, INCA integrates remote sensing data with surface station observations, combining their respective strengths to provide accurate and comprehensive weather forecasts. Specifically, the precipitation analysis within INCA utilises interpolated station data with from ground-based radar data. This dual approach leverages the precise point measurements from surface stations and the spatial coverage provided by radar, capturing precipitating cells that may not be detected by station measurements alone. The system aggregates 1-minute precipitation data from a network of automated stations into 5-minute intervals, interpolating these values onto a 1×1 km grid using distance weighting methods. To ensure the accuracy of widespread precipitation detection, only the nearest eight stations are considered. The radar data, available at 5-minute intervals, are bilinearly interpolated onto the INCA grid and then scaled to correct for range-dependent biases and topographic shielding effects. This process involves a climatological scaling factor for each month and a subsequent real-time adjustment based on station observations. The system allows for a spatial shift to account for factors such as wind drift, aligning radar data more accurately with surface observations. These two fields are integrated to provide a more comprehensive final precipitation analysis. 

CAPE is estimated using the 3D temperature corrected fields from the INCA system, as described by \citet{steinheimer2007improved}. Unlike the high temporal resolution of INCA\_a, CAPE is available at 1-hour temporal resolution. Among the CAPE values estimated by INCA analysis model runs, only one estimation from the closest time step to the prediction time is given to the model as input data.

\begin{figure}
    \includegraphics[width=\textwidth]{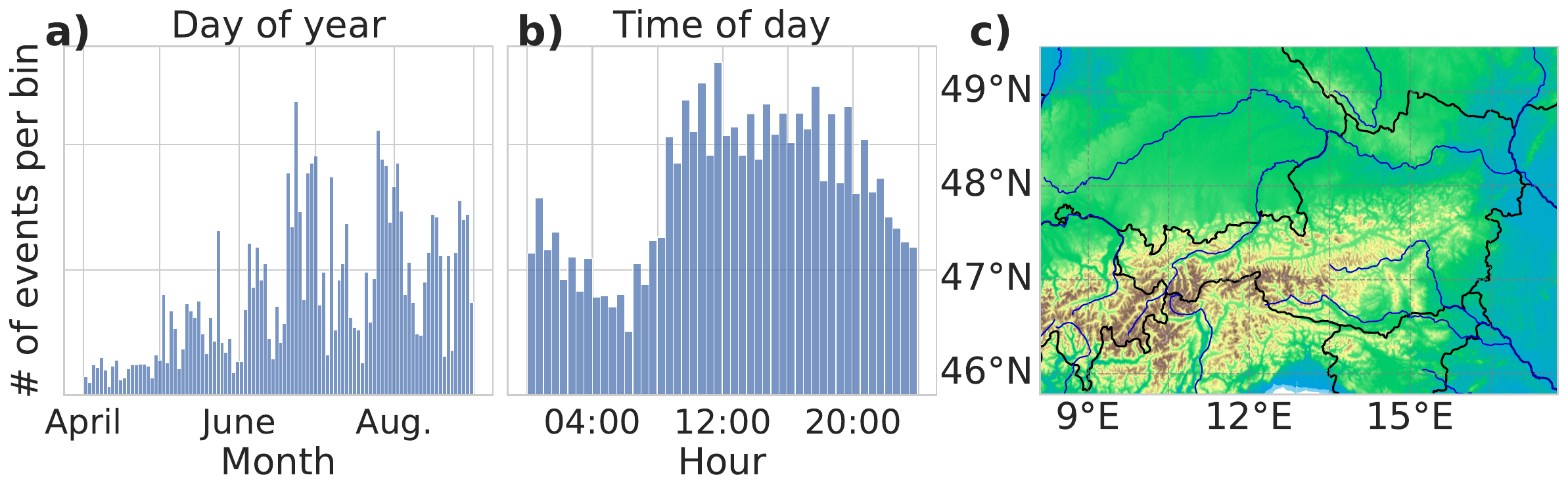}
    \caption{Summary of the dataset in temporal distribution of the events a) per day of year, b) time of day, and c) spatial extent of the study domain.}
    \label{fig:dataSummary}
\end{figure}

\section{Methodology}

\subsection{Data sampling and preprocessing}\label{ssec:Preprocess}

The dataset contains more than 7800 events, each spanning over 4 hours to support both this study and future research, sampled over 5 years, from 2019 to 2023, during summer periods from April to September, when convective activity is highest over Austria. Temporal sampling of the events is stratified in order to avoid clear-sky conditions dominating the dataset and to increase the focus on convective events. Assuming the second half of INCA precipitation values of each event is set as the target, we used spatially-averaged precipitation values of each possible time step to compute the standard deviation over consecutive periods of two hours across the temporal domain. We then sampled the first time step to be predicted from these time stamps by using an inverse probability weighting of these temporal standard deviations for stratification, thus increasing the probability of occurrence of precipitation fields with rapid growth/decay. This resulted in more events occurring from mid-June to mid-August (Figure \ref{fig:dataSummary}a), together with a diurnal pattern showing more events during the daytime (Figure \ref{fig:dataSummary}b), suggesting the stratification was useful in obtaining a larger number of convective events in the dataset.

The spatial domain of the dataset is the same as the operational rapid-INCA domain (shown in Figure \ref{fig:dataSummary}c). This region is characterised by complex topography, including the Eastern Alps, which significantly influences weather patterns and poses challenges for meteorological modelling. The area experiences frequent convective storms, particularly in the summer, which are often driven by the interaction of moist air masses with the mountainous terrain. These storms can lead to heavy precipitation, lightning, and other severe weather phenomena, making accurate forecasting in this region particularly important.

We considered the 6438 events from the first four years of convective seasons as the training dataset and kept the 1395 events from the last season for validation and testing. Among the 2023 data, we used the 221 events from Mondays \cite[parallel to the approach in][]{Ravuri2021} as the validation dataset and used the remaining as the test dataset. In order to address the skewed distribution of precipitation estimates from radar observations and INCA analysis, we followed the transformation approach of \citet{Leinonen2023b} and used the following function:
\begin{equation}
	f(R) = \begin{cases} 
		\log_{10} R & \text{if } R \geq 0.1 \, \text{mm h}^{-1} \\
		\log_{10} 0.02 & \text{if } R < 0.1 \, \text{mm h}^{-1}
	\end{cases}
\end{equation}

Other variables were transformed by using the mean ($\mu$) and standard deviation ($\sigma$) of each variable computed over the training dataset as
\begin{equation}
	z = \frac{x - \mu}{\sigma}
\end{equation}

Last but not least, all data that were not already in the MGI / Austria Lambert (EPSG code 31287) projection system were reprojected to this projection using GDAL \citep{GDAL}.

\subsection{Modified Transformers architecture}

We adapted the Earthformer satellite-to-radar nowcasting model (EF-Sat2Rad) architecture from \citet{Kucuk2024}, which is a modified version of the Earthformer package \citep{Earthformer}. Earthformer is a space-time Transformer designed to address problems in Earth system science by leveraging the strength of Transformers \citep{Vaswani2017} to compute space-time attention with a comprehensive patching framework for spatiotemporal data. Utilising a U-Net style architecture for patching the data to compute attention cuboids, the model is able to extract information across scales, making it particularly powerful for learning environmental processes that operate on different scales while interacting with each other.

The objective of the model is to predict precipitation intensity for the coming 90 minutes over the operational INCA domain with 1 km and 5 minute resolutions in space and time, given observation data from space- and ground-based sensors, NWP and analysis products, merged with other geospatial information provided to the model as input spanning the past 95 minutes from the prediction time. 

In addition to the spatiotemporal data with 5-minute temporal resolution described in Section \ref{sec:Data}, we also used auxiliary variables that are static or with coarser temporal resolution, namely CAPE, elevation, time of the events, and coordinates of the domain. Indicating the amount of potential energy for convective activity, CAPE is commonly used as a diagnostic for atmospheric instability. Elevation data 
are also used to support the model in learning orography-related effects. In addition, the day of the year and the hour of the day information of the first lead time to be predicted are used together with the coordinates of the domain to provide further information about the general spatial climatology of the domain. Significant benefit from these parameters is not expected; though they may help the model to generalise better to unseen spatiotemporal domains in future applications.

The training strategy was largely similar to that of \citet{Kucuk2024}, including the use of mean squared error (MSE) as the loss function, which is inversely scaled with increasing lead time to force the model to be more accurate in the earlier lead times. However, given the increased complexity and larger amount of input data, training was more unstable than the earlier model, leading us to set the maximum learning rate to $3 \times 10^{-4}$ for the one cycle learning rate scheduler and add L2 regularization to the loss values. An early stopping framework was utilised to avoid overfitting, which stopped the optimisation in the 41st epoch. We refer to the code repository accompanying this article for further details of the model configuration. Overall, the trained model had 12.3 million parameters and required around 2.5 hours on a virtual machine with a graphical processing unit (GPU) of 48 GB memory on an RTX A6000 graphics card. 

\subsection{INCA operational predictions}
The INCA precipitation (INCA\_p) nowcasting system includes both observation-based vectors, using motion derived from previous radar fields, and NWP model wind vectors. This ensures a smooth transition between nowcasting and NWP displacement and maintains accuracy over time. Spurious correlations obtained from comparing radar fields, which could imply unrealistically large motion speeds, are filtered meteorologically by comparison with AROME \citep{Auger2015} wind fields at 500 and 700 hPa. This comprehensive approach significantly enhances the reliability of precipitation forecasts, particularly in complex terrains. The nowcasting uses these final extrapolation vectors to advect the latest INCA precipitation analysis using Lagrangian advection \citep{haiden2011integrated}. The growth and decay of the field are not accounted for, and consequently, the system can be described as variance-preserving.

\subsection{Pysteps predictions}
Pysteps is an open-source Python implementation of the short-term ensemble prediction system \citep[STEPS;][]{Bowler2006}, initially developed by \citet{Pulkkinen2019}. It is an initiative for probabilistic nowcasting that is under continuous development, incorporating new features and advancements as described by \citet{Imhoff2023}.

Within the Pysteps initiative, we use the spectral prognosis method \citep[S-PROG;][]{Seed2003}, which smooths the field by removing unpredictable small scales using an auto-regressive filter. In addition, a probability matching method is applied as a post-processing step to preserve the distribution of values. Consequently, this is still a variance-preserving system. However, it has different features from the INCA operational system. 

INCA analysis data are used as the Pysteps input after being log-transformed using the same approach for EF4INCA, described in Section \ref{ssec:Preprocess}. Source code to replicate the Pysteps predictions is available in the code repository.

\section[Results]{Results}
\subsection{Model performance}

\begin{figure}[h]
	\includegraphics[width=\textwidth]{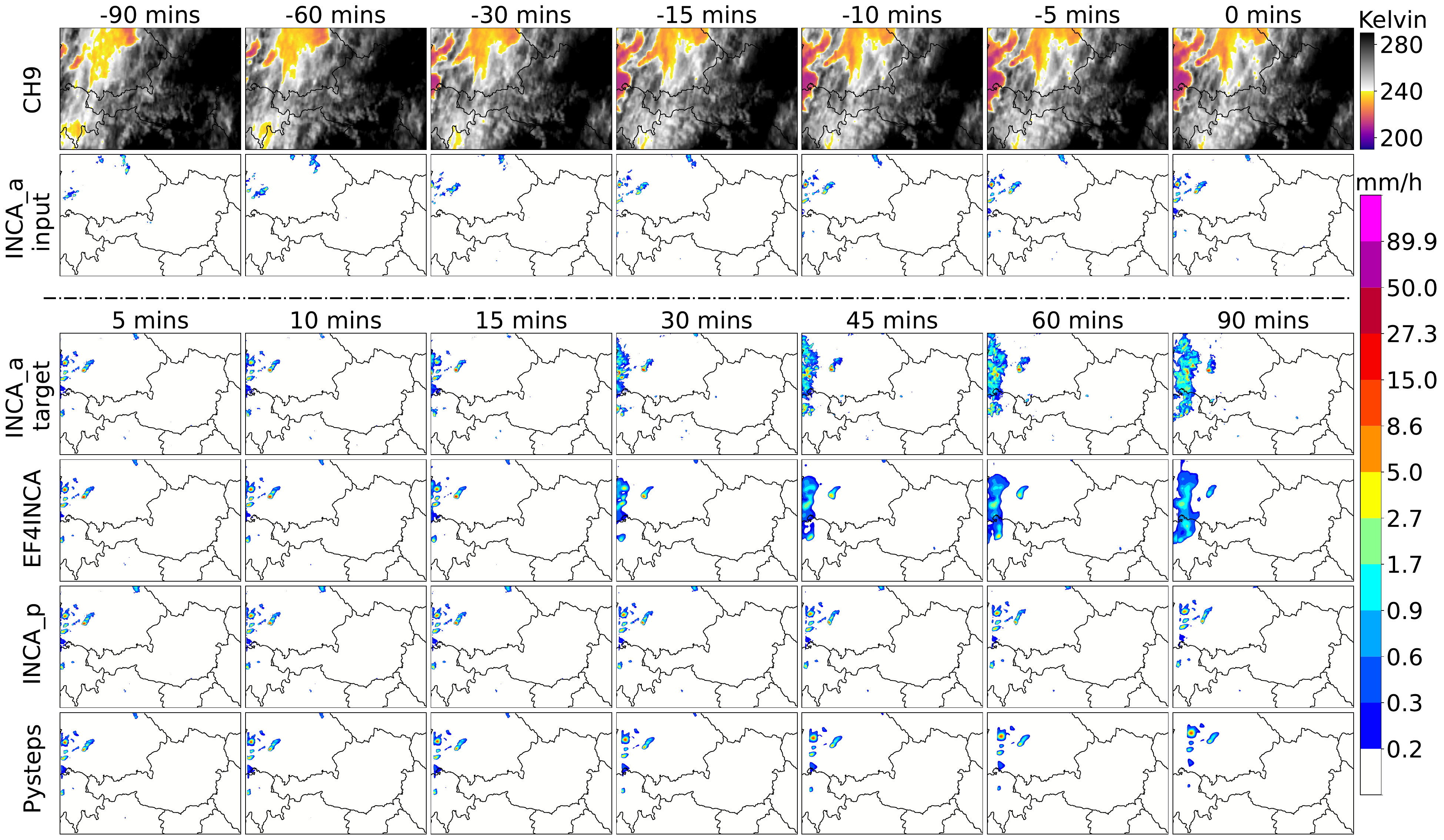}	
	\caption{Model output for a case study on a convective event with the predictions delivered for 5 May 2023 14:55 UTC onwards. First two rows present the most distinguished input channels as CH9 ($10.8 \mu m$) and INCA analysis data (INCA\_a), while rows after the dashed line present the target as future INCA\_a values, followed by predictions from EF4INCA, INCA (INCA\_p), and Pysteps, respectively. See Figure \ref{fig:modelOutputExtended} for the extended version presenting all of the input data.}
	\label{fig:modelOutput}
\end{figure}

We present model performance first visually through a case study and then through prediction accuracy metrics quantified over the complete test dataset. A selected case covering initiation of a convective cell is shown in Figure \ref{fig:modelOutput}, where the first two rows show the most distinctive input channels for the EF4INCA model as CH9 from SEVIRI and observed INCA analysis (INCA\_a) estimations, together with the target variable and predictions from EF4INCA, INCA nowcasting prediction (INCA\_p), and Pysteps models, respectively (See Figure \ref{fig:modelOutputExtended} for the extended version of the figure, as well as further examples in the repository containing model weights for hand-picked and randomly selected events visualising good and bad performance). 

Until the prediction time of the case shown in Figure \ref{fig:modelOutput}, INCA\_a input data provide almost no information on the upcoming event since data streams for that product have not observed any precipitation. However, satellite observations indicate a sharp temperature gradient across space, suggesting atmospheric instability that can lead to convective activity. EF4INCA model successfully reproduced the rapidly growing spatial extent of the convective activity through the lead times, presumably by using the diagnostics available in satellite data. However, INCA\_p and Pysteps did not use the information in satellite data efficiently and only extrapolated the small precipitation field observed before the prediction time, thus failed to predict the growth of this convective activity.

Alongside the rapidly growing convective fields, EF4INCA outperformed INCA\_p and Pysteps in other challenging cases for nowcasting with complex temporal features such as rapid movement and decay of precipitation fields (see the repository containing model weights for further case studies). However, EF4INCA predictions lack sharp structure in the predicted fields, particularly in the later lead times, while the other, variance-preserving, methods maintain the sharp structure of predicted precipitation fields throughout predicted time steps. 

\begin{figure}
	\includegraphics[width=\textwidth]{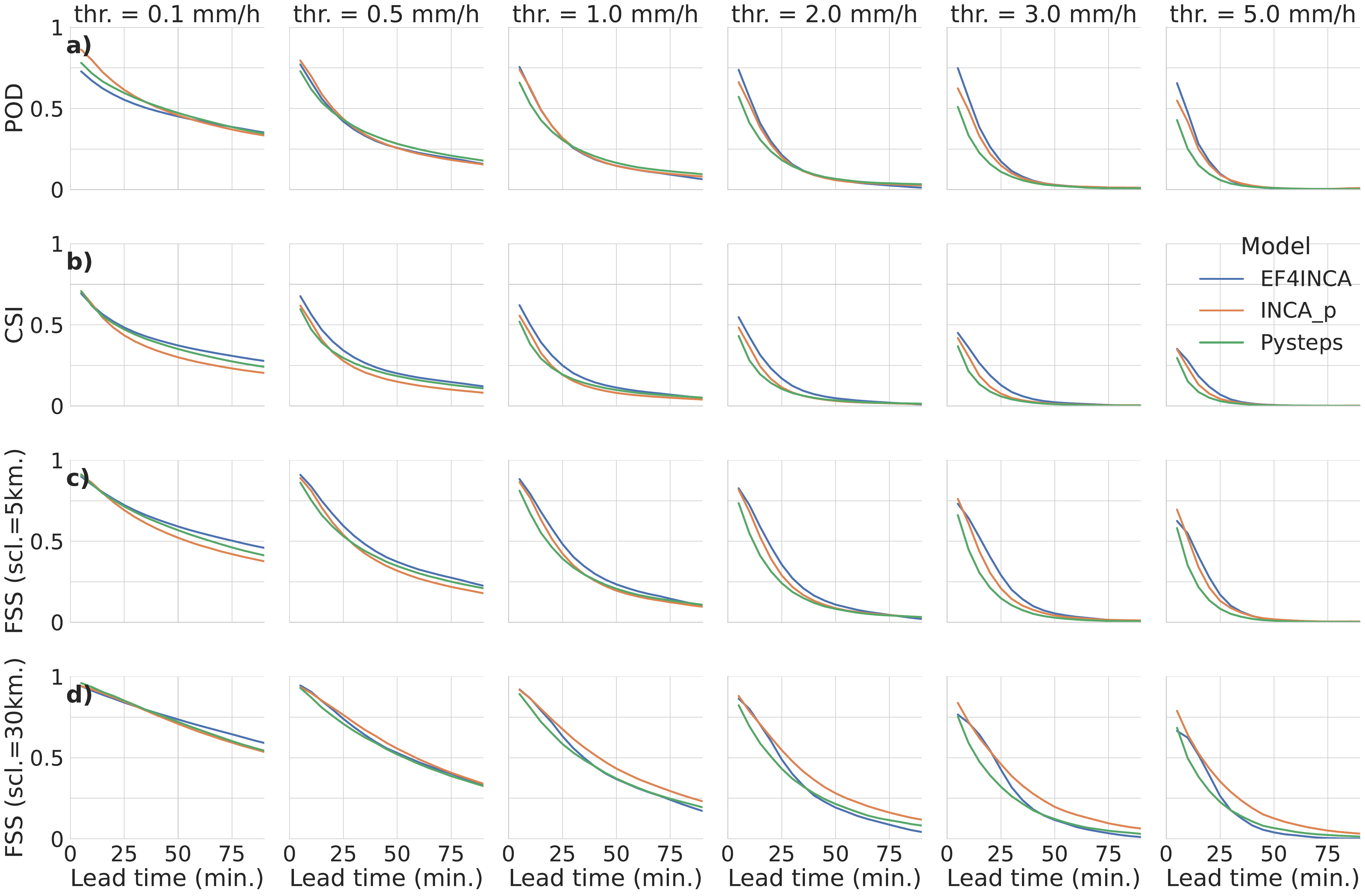}
	\caption{Deterministic verification scores of the models computed over the entire test dataset: a) Critical success index (CSI), b) Probability of detection (POD), and fractional skill score (FSS) computed with spatial scales of c) 5 and d) 30 kilometres.}
	\label{fig:modelPerf}
\end{figure}

In addition to the case studies, we quantified model performance using various skill scores computed over the whole test dataset as well. The probability of detection \cite[POD;][]{Hogan2011}, also known as hit rate, scores for various thresholds are summarised in Figure \ref{fig:modelPerf}a. While the variance-preserving models received higher POD scores in thresholds smaller than 1 mm/h, EF4INCA scored higher in larger thresholds. 
Critical success index \cite[CSI;][]{Hogan2011}, also known as the threat score, was also computed in order to consider event detection and false alarms simultaneously. Unlike POD, EF4INCA received higher CSI scores than the other two models across all thresholds (Figure \ref{fig:modelPerf}b) This suggests superior performance of EF4INCA in predicting location of the precipitation fields without delivering too many false alarms, particularly in low precipitation values. 

CSI and POD are computed at grid cell level, which heavily penalises location inaccuracies. Therefore, we also computed the fractional skill score \cite[FSS;][]{Roberts2008}, which accounts for neighbouring grid cells when computing prediction skill. Using a spatial scale of 5 km, EF4INCA has the highest FSS for all thresholds (Figure \ref{fig:modelPerf}c), indicating higher performance than the variance-preserving models when location precision is taken into account. However, at larger spatial scales like 30 km, INCA\_p achieves higher FSS than the other models (Figure \ref{fig:modelPerf}d), suggesting better amplitude predictions than EF4INCA at lower precision in location. 

\subsection{Model interpretation}

Permutation tests were performed on the input data to gain insights into the EF4INCA model to improve trustworthiness of the model and to understand the most skilful input channels for guiding future developments. Specifically, we permuted one input channel of the test dataset at a time, fed the modified data for inference and finally computed FSS with a 5 km kernel size. Relative importance (RI) of any permuted input data stream $X$, as $RI_X$, was estimated as

\begin{equation}\label{eq:skill}
	RI_X = 1 - FSS_X / FSS_{ref}
\end{equation}

where $FSS_X$ is the skill score of the modified data and $FSS_{ref}$ is the reference FSS of the original test dataset presented in Figure \ref{fig:modelPerf}c. In this context, a lower $FSS_X$ compared to $FSS_{ref}$ indicates a greater negative impact of the permutation, highlighting the importance of the permuted input. In order to compensate for the reduced values in skill scores, we normalised the RI values per threshold and prediction time step, which increased interpretability. 

Normalised RI values of the spatiotemporal input data are shown in Figure \ref{fig:modelInterpret} (see Figure \ref{fig:modelInterpretAux} in the appendix for auxiliary input data). INCA\_a contains the highest RI, particularly for the early lead times and low precipitation thresholds, which is intuitive since the prediction target in early lead times is expected to be similar to its previous time steps. Radar data have the second highest RI for all thresholds with an increase in RI from lead time 5 to 15, presumably due to complementing INCA\_a with a slight latency. This increase is then followed by a reduction in RI, parallel to that of INCA\_a.

\begin{figure}
	\includegraphics[width=\textwidth]{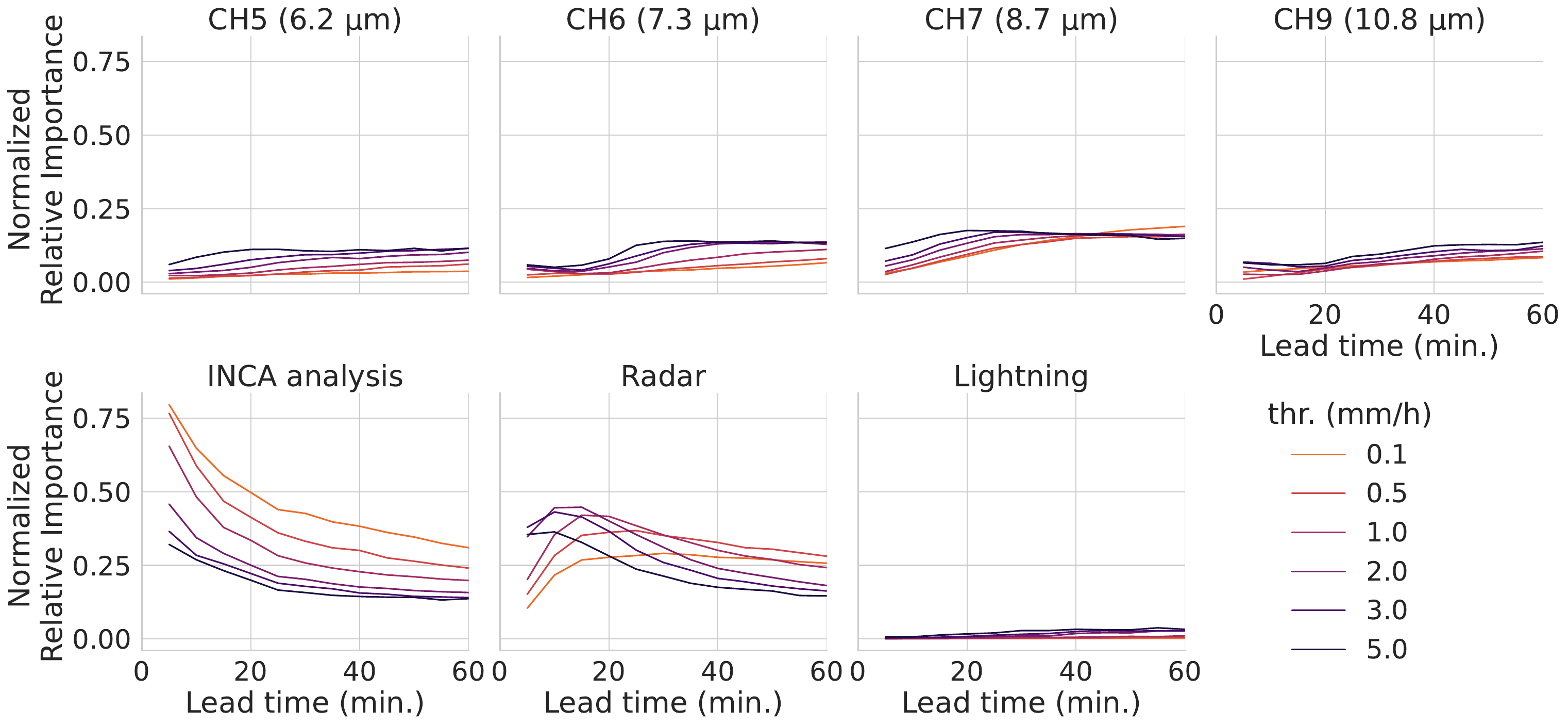}
	\caption{Normalized relative importance of input channels for model performance computed using Equation \ref{eq:skill}, normalised per lead time and permuted channel.}
	\label{fig:modelInterpret}
\end{figure}

The RI of satellite observations is less than precipitation products, though, interestingly, it increases with increasing precipitation threshold and extended lead times. This is expected since it is known that satellite meteorology can detect convective activity through overshooting tops earlier than other ground-based observations \citep{Dworak2012}. Among the MSG channels, CH7 has the highest RI, closely followed by CH9 and CH6. High RI of CH7 is expected since the wavelength it is centred is sensitive to cloud top phase \citep{Holmlund2021,Schmit2017}, providing valuable information about cloud microphysics. Model interpretation of the RI of CH9 also corroborates the literature, since it has been shown to be a useful diagnostic for severe weather in studies with different methodologies \citep{Velden2006, Hilburn2020, Kucuk2024}. Finally, RI in the water vapour bands CH6 and CH5 is also intuitive since they contain information about water vapour content in the atmosphere, which supports monitoring instability and humidity in the atmosphere \citep{Holmlund2021}, and are used to track evolution of synoptic-scale atmospheric phenomena \citep{Line2016}. However, these channels provide the mean temperature of their targeted layer in the atmosphere, which makes them prone to noise \citep{Schmit2018}.

Finally, we found the lowest RI within the spatiotemporal input data streams for lightning data, which is surprising since it is known that lightning activity contains skilful information for detecting severe weather \citep{Deierling2008, Kucuk2024}. While the slight increase in RI of lightning data with higher precipitation thresholds agrees with the literature, the overall low RI compared to the other predictors may have multiple reasons such as (i) the model's focus on the other data streams, particularly precipitation products, (ii) initial filtration of the detected lightning events being too aggressive, and (iii) discrepancies between ground- and space-based lightning detection systems. Analysis of the RI in lightning data obtained from different data streams such as the Lightning Imager of the Meteosat Third Generation satellite, or different preprocessing of ground-based lightning detection remains an open question for future studies.

\section[Discussion]{Discussion}
Leveraging the attention mechanism \citep{Vaswani2017} with an efficient spatiotemporal tokenizing approach and a hierarchical encoder-decoder architecture to learn spatiotemporal interactions across scales, Earthformer \citep{Earthformer} based models offer powerful solutions for Earth system science problems. As a modified version of the Earthformer model, the EF4INCA model is capable of exploiting different data streams for precipitation nowcasting, which is particularly important for developing advanced nowcasting models integrating data from various channels.

Early detection of convective activity is a key task in nowcasting since convective storms can trigger extreme events such as heavy precipitation. As ground-based observations and conventional NWP models do not provide sufficient information about such cases, it is critical to integrate space-based observations in prediction systems efficiently. Being able to extract information across data streams, the EF4INCA model outperforms conventional models in predicting the onset of convection and its further development by using information available in satellite-based observations. On the contrary, conventional models have limited capacity in nowcasting onset of convection since they do not use satellite data that may provide diagnostics for initiation of convection but only use the observed precipitation fields large enough to advect.

Another point where EF4INCA excels is in the location accuracy of the predicted fields. Conventional models compute motion vectors of fields for advecting observations, which is prone to errors, particularly in cases with complex temporal features in rapidly moving, growing, or decaying fields. Thanks to its efficiency in learning these features, the EF4INCA model can predict the movement of precipitation fields together with their temporal development, providing higher accuracy in precipitation location than conventional models.

Regarding the structure of predicted fields, EF4INCA performs worse than the conventional models since they are designed to preserve variance. As a deterministic model with an error-minimising objective, EF4INCA has a tendency to predict smoothed fields to satisfy its objective, resulting in blurred predictions where the model reflects high prediction uncertainty. This trade-off between location and structure of predicted fields is also evident in the FSS of the models, where higher scores are obtained by EF4INCA at small spatial scales while at large spatial scales, INCA\_p scores higher.

Finally, we discuss the runtime and computational resources required for running these models operationally. Despite the high initial cost during training, machine learning models are orders of magnitude cheaper for inference compared to conventional, physics-based models, creating novel opportunities for expanding applications in different domains, particularly in ensemble generation. EF4INCA can make one prediction, using a GPU with 12 GB RAM, in around 0.025 seconds, while it takes close to 2 minutes for INCA\_p using a single CPU. Given the structural differences between the hardware used by the models, direct comparison of their runtime may be imprecise. Nevertheless, such a significant gain in runtime by using a modern model on a consumer-grade GPU is both feasible and appealing.

\section[Conclusions]{Conclusions}

In this study, we presented EF4INCA, a new precipitation nowcasting model that efficiently integrates data from different streams by leveraging a Transformer-based architecture. EF4INCA demonstrates strong performance in predicting the location and shape of precipitation fields, particularly in challenging cases with temporally complex features such as the onset of convection, as well as rapidly-moving and decaying fields. It outperforms conventional nowcasting models such as INCA\_p in most of the skill scores. However, as a deterministic and error-minimising model, EF4INCA has the tendency to generate overly smooth fields at later time steps of the predictions. The smoothing effect, while beneficial for overall accuracy, results in loss of fine-scale structural details of the predicted precipitation fields at longer lead times compared to the conventional models. This trade-off highlights the model's strength in capturing large-scale patterns and movements, while also pointing to areas for future improvement in preserving small-scale precipitation structures.

EF4INCA was interpreted through permutation experiments to gain further insights from the model and identify the most skilful input data streams. Among the input data, we found precipitation products to contain the highest RI, followed by SEVIRI infrared channels CH7 and CH9. These intuitive results, also corroborating the literature, increase trust in the model. 

The high RI of INCA precipitation analysis and radar data underscores the crucial role of physics-based model outputs and ground-based observations in achieving comprehensive and accurate forecasts. In addition to the above findings, our analysis showed that satellite data such as CH7 and CH9 channels provide valuable information on cloud microphysics and severe weather diagnostics, especially during the early phases of convective activity where precipitation observations have limited information capacity. Therefore, integration of these diverse data sources is essential for improving prediction accuracy, especially in regions with complex topography and rapidly changing weather conditions.

In order to facilitate further development, we have shared the code repository and trained weights of the model, together with the dataset prepared for the model. Future studies can use this as a benchmark for developing local area models integrating observation data with analysis products. Future research should focus on improving the tokenization and downscaling of the Transformer backbone of the architecture. Additionally, developing methods for probabilistic predictions with sound uncertainty quantification will be crucial for enhancing the model's reliability and usefulness for complex hazard management frameworks. Finally, in-depth exploration of the RI of lightning observations from different data streams remains a topic for future studies. Such advancements could pave the way for more accurate weather forecasts with early detection capabilities, benefiting a wide range of meteorological applications. Ultimately, the EF4INCA model and the insights gained from this study represent a significant step forward in integrated precipitation nowcasting with ML, offering improved predictions for complex weather phenomena and laying the groundwork for future innovations in data-driven weather predictions.
	
\bibliography{ms}
	
\newpage

\paragraph{Acknowledgments}
We are grateful for the assistance of Vera Meyer and Lukas T\"uchler for the initial access to raw radar and lightning data. 

\paragraph{Funding Statement}
\c{C}.K. is supported by the European Organisation for the Exploitation of Meteorological Satellites (EUMETSAT) fellowship ``FUSEDCAST: A fused data approach to the nowcasting of severe weather''

\paragraph{Competing Interests}
The authors declare none.

\paragraph{Data Availability Statement}
Data curated for this study is available in \url{https://doi.org/10.5281/zenodo.13740314}. 

Code repository of the EF4INCA model is available in \url{https://github.com/caglarkucuk/earthformer-multisource-to-inca}. 

EF4INCA model weights and figures of further case studies are available in \url{https://doi.org/10.5281/zenodo.13768228}.


\paragraph{Author Contributions}
Conceptualization: \c{C}.K., A.A., M.D.; data curation, formal analysis, methodology, software, visualization, and writing -- original draft: \c{C}.K.; verification: \c{C}.K., A.A.; writing -- review \& editing: \c{C}.K., A.A., M.D.

\newpage

\appendix
 
 \section{Extended version of Figure 2}
 \begin{figure}[h]
 	\includegraphics[width=\textwidth]{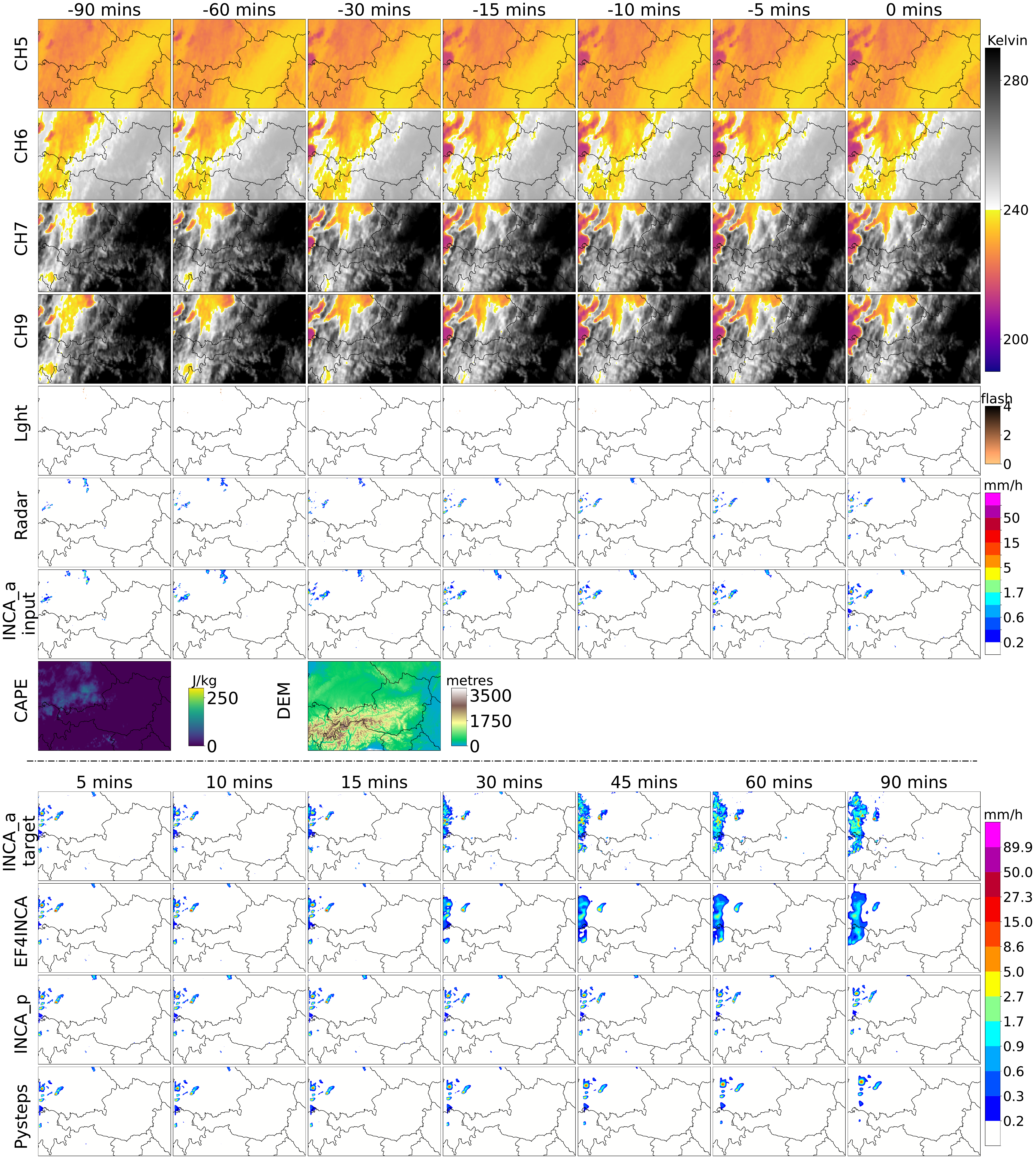}	
 	\caption{Extended version of Figure \ref{fig:modelOutput}. Model output for a case study on the onset of a convective event with the predictions delivered for 5 May 2023 14:55 UTC onwards. First four rows present input data from SEVIRI for channels CH5 ($6.2 \mu m$), CH6 ($7.3 \mu m$), CH7 ($8.7 \mu m$), and CH9 ($10.8 \mu m$), respectively. Fifth row is lightning flash counts. Sixth and seventh rows present precipitation estimates from radar observations and INCA analysis, respectively. Eighth row contains CAPE estimates from INCA analysis and digital elevation model (DEM) of the study domain. First row after the dashed line shows the target precipitation field to be predicted, followed by predictions from EF4INCA, INCA, and Pysteps models, respectively.}
 	\label{fig:modelOutputExtended}
 \end{figure}

\newpage
\section{Model interpretation for auxiliary input data}
    
\begin{figure}[h]
    \centering
    \includegraphics[width=.75\textwidth]{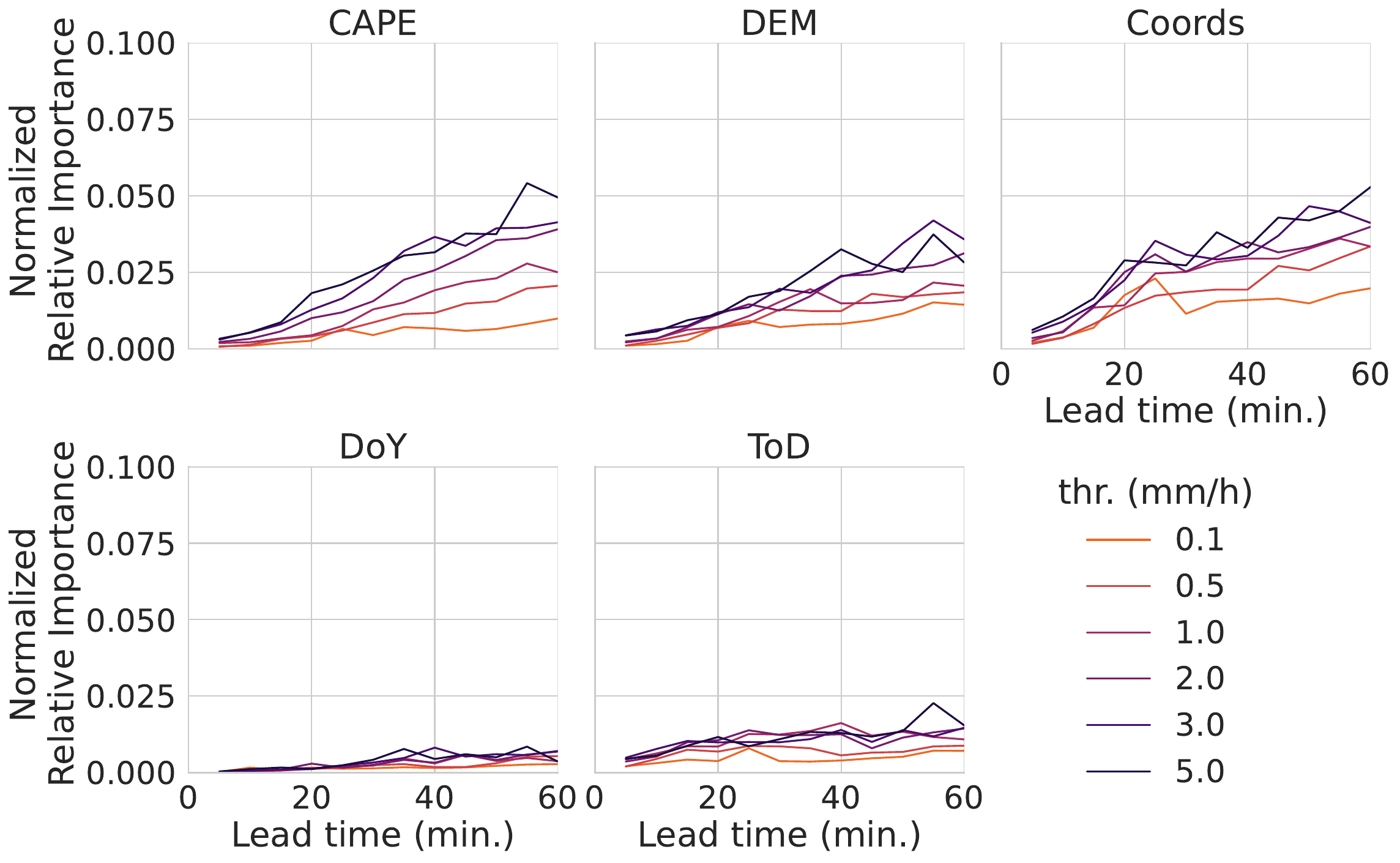}
    \caption{Same as Figure \ref{fig:modelInterpret}, but for auxiliary variables with non-homogeneous temporal resolution, with different scaling of y-axis for better interpretability.}
    \label{fig:modelInterpretAux}
\end{figure}
	
\end{document}